\documentstyle[12pt,epsf,psfig]{article}
\def \beq{\begin{equation}}
\def \eeq{\end{equation}}
\def \beqar{\begin{eqnarray}}
\def \eeqar{\end{eqnarray}}

\begin{document}
\bigskip
\bigskip
\makebox[2cm]{}\\[-0.5in]
\begin{flushright}
\begin{tabular}{l}
UFTP 431/1996 \\ TUM/T39-96-29
\end{tabular}
\end{flushright}
\vskip0.4cm
\begin{center}
  {\bf Phenomenology of IR-renormalons in inclusive processes}\footnote{Work
    supported in part by BMBF}\\
\end{center}
\bigskip
\centerline{\it M.~Maul, E.~Stein, A.~Sch\"afer}
\centerline{\it Institute for Theoretical Physics, 
Johann Wolfgang Goethe-Universit\"at,}
\centerline{\it D-60054 Frankfurt am Main, Germany} 
\medskip
\centerline{and}
\medskip
\centerline{\it L. Mankiewicz\footnote{On leave of absence from N. Copernicus
  Astronomical Center, Polish Academy of Science, ul. Bartycka 18, PL--00-716
  Warsaw (Poland)}}
\centerline{\it Institute for Theoretical Physics} 
\centerline{\it TU-M\"unchen, D-85747 Garching, Germany} \medskip
\begin{quote}
  We have compared the existing experimental data on the leading power
  corrections to the structure functions $F_2(x,Q^2)$, $F_3(x,Q^2)$, and
  $F_L(x,Q^2)$ with the IR-renormalon model predictions for higher-twist
  contributions. Our analysis shows that the model properly describes the
  $x$-dependence, but typically falls short by a factor 2 or 3 as far as the
  magnitude of higher twist corrections is concerned.
\end{quote}
\bigskip 

\newpage

Measurements of QCD observables now have reached such high precision that power
corrections to the structure functions can often be extracted with a reasonable
accuracy from the existing data. The situation on the theoretical side is much
less clear. In the best understood case of deep inelastic scattering, the
relevant contributions can be attributed in the framework of operator product
expansion (OPE) to matrix elements of higher twist operators \cite{hightwist},
but their determination in QCD is ambiguous due to occurrence of power
divergences \cite{Mart96}. From the phenomenological point of view, however,
attempts to compute these matrix elements using e.g., QCD sum rules or the bag
model have provided results which seem to have at least the right order of
magnitude \cite{BBK,Stein,Braun95,Bag} as compared with available experimental
estimates.

Recently there has been some interest in another phenomenological approach to
power-suppressed corrections in QCD \cite{renormalons}, based on the fact that
the only possibility to interpret the higher-order radiative corrections in a
consistent manner (i.e. as asymptotic series) requires the existence of power
suppressed terms \cite{Beneke}.  The IR-renormalon contributions occur
because certain classes of higher order radiative corrections to twist-2 are
sensitive to large distances, contrary to the spirit of OPE. Although these
divergences have to cancel with UV power divergencies in matrix elements of
twist-4 operators i.e., they are totally spurious, there are arguments which
suggest that they reflect the correct shape, if not the magnitude, of the
power-suppressed terms \cite{DMW95,Nord96,BBM97}.  In this sense we define the
prediction of the renormalon model of higher-twist corrections to a DIS
structure function as the power-suppressed uncertainty which occurs in the
perturbative expansion of the Wilson coefficient to the twist-2 contribution.

Commonly the divergent series of radiative corrections is regarded as an
asymptotic series and defined by its Borel integral. The actual calculations
are done in a large-$N_F$ limit which allows to resum the fermion
bubble-chain to all orders yielding the coefficient of the $\alpha_S^n
N_F^{n-1}$ - term exactly. Subsequently, it is converted into the exact
coefficient of the $\alpha_S^n \beta_0^{n-1}$ - term by the substitution $N_F
\to N_F - 33/2 = - 6 \pi \beta_0$, known as the 'Naive Non-Abelianization'
(NNA) (see the last two references of \cite{Beneke}). The asymptotic character
of the resulting perturbative series leads to resummation ambiguities - a
singularity in the Borel integral destroys the unambiguous reconstruction of
the series and shows up as a factorial divergence of the coefficients of the
perturbative expansion.  The general uncertainty in the perturbative prediction
can be estimated to be of the order of the minimal term in the expansion, or by
taking the imaginary part (divided by $\pi$) of the Borel integral. Both
procedures lead to resummation ambiguities of the form $C\times
\left(\Lambda^2/Q^2\right)^r$, with $r=1$ for the leading IR renormalon, and a
numerical coefficient $C$ which either can be taken as it comes out from the
NNA calculation, see below, or can be fitted as a free parameter.

The unpolarized hadronic scattering tensor for leptons scattering 
off nucleons can be divided into three structure functions  \cite{BBDM78}
\beqar
\label{10}
W_{\mu \nu}(p,q) &=& 
\left(g_{\mu \nu} - \frac{q_\mu q_\nu}{q^2} \right) 
\frac{F_L(x,Q^2)}{2x}
            + \left( - \frac{p_\mu p_\nu}{(p \cdot q)^2}q^2  
            +  \frac{p_\mu q_\nu + p_\nu q_\mu}{p \cdot q} - g_{\mu \nu} \right
) \frac{F_2(x,Q^2)}{2x}
\nonumber \\ && - i\epsilon_{\mu \nu \alpha \beta}\frac{p^\alpha q^\beta}{p 
\cdot q} 
                F_3 (x,Q^2) \quad.
\eeqar            
Here $x = Q^2/(2 p \cdot q)$ and $Q^2 = -q^2$.  The structure functions $F_i,
i=L,2,3$ can be generally decomposed in the following manner
\beq
\label{20}
F_i( x,Q^2) = F_i^{\rm t-2}(x,Q^2) + \frac{1}{Q^2} h_i^{\rm TMC}(x,Q^2)
                                + \frac{1}{Q^2} h_i(x,Q^2) 
                                + {\cal O} ( \frac{1}{Q^4})\quad,
\eeq
where $F_i^{\rm t-2}$ describes the leading twist-2 contribution.  $h_i^{\rm
  TMC}$ describes the target mass corrections which are directly related to
twist-2 matrix elements \cite{FG80}.  $h_i$ contains the genuine twist-4
contribution which is in principle sensitive to multiparton correlations within
the hadron, and which we want to estimate using the renormalon model. Note that
$h_i$ has dimension 2 (if radiative corrections are neglected) and hence it has
to be proportional to a certain mass scale $\mu^2$. In the analysis of the
experimental data used in this paper the coefficients $h_i$ were extracted by
accounting for target mass corrections up to order ${\cal O}(M_N^4/Q^4)$,
except for the case of $h_{\rm L}$ where target mass corrections have been
accounted for to the order ${\cal O}(M_N^2/Q^2)$ only. In the analysis of the
twist-4 contributions \cite{NMC92,VM92,Si96} it is common to neglect the $Q^2$
dependence of $h_i$, which is due to radiative corrections, and to sum target
mass corrections and the twist-2 contributions into a leading-twist (LT)
structure function thus arriving at the notation
\beq
\label{30}
F_i( x,Q^2) = F_i^{(LT)}(x,Q^2)+  \frac{1}{Q^2} h_i(x) 
            = F_i^{(LT)}(x,Q^2) \left( 1 + \frac{C_i(x)}{Q^2} \right)
\quad.
\eeq
\bigskip

Now we shall shortly summarize the main features of the model. In the
renormalon approach the expression for the higher twist correction $h_i(x)$ 
to a
structure function $F_i$ has the form of a Mellin convolution \cite{SMMS96} 
\beq
h_i = P_i \otimes F_i^{\rm t-2} 
\eeq 
of a coefficient $P_i(z)$, and the twist-2 part $F_i^{\rm t-2}(x)$ of the
structure function $F_i$. The mass scale $\mu^2$ which enters $P_i(z)$ equals
to $\mu^2_R = 8 \pi C_F \Lambda_s^2 e^{-C_s}/\beta_0$ where $C_F = 4/3$,
$\Lambda_{\rm s}$ is the QCD scale parameter in a given renormalization scheme
s, and $C_{\rm s}$ is determined by the finite part of the fermion loop,
$C_{\overline{\rm MS}} = -5/3$. This identification, which follows from the NNA
prescription, is RG invariant only in the strict large-$N_f$ limit. Another
difficulty stems from the observation that within the logic of the model one
cannot distinguish between LO and NLO parametrizations of twist-2 structure
functions $F_i^{\rm t-2}(x)$. In the following we will consequently use the LO
GRV parametrizations of parton distributions \cite{GRV94}, but we have checked
that qualitatively the same conclusions follow from a calculation which
incorporates the NLO parametrizations. In the present analysis we always refer
to a region of $Q^2$ around $2$ GeV$^2$ i.e., below the charm threshold, and
accordingly we adopt the $N_F = 3$ value of the $\Lambda$ parameter from the LO
GRV fit, $\Lambda = 232$ MeV. We mention also that due to the two possible
contour deformations above or below the pole in the Borel integral, as
discussed above, in principle the overall sign of the coefficients $P_i(z)$
cannot be uniquely determined.

We shall state from the very beginning that because the IR-renormalon result is
proportional to the twist-2 contribution, it cannot be expected to describe the
complete twist-4 correction which contains a genuine multiple field
correlation, and which depends therefore on the exact internal hadron wave
function.  This distinction has a real physical meaning as can be seen from the
following argument. Let us assume that the same structure functions $F(x,Q^2)$
are measured for different hadrons and that the lowest-order (in $\alpha_s$)
leading twist ($F^{\rm t-2}(x)$) and $1/Q^2$ parts ($F^{\rm t-4}(x)$) can be
separated experimentally.  
Then the renormalon contribution cancels for the
difference of the ratio of moments
\begin{equation}
\left( {M_n(F^{\rm t-4})\over M_n(F^{\rm t-2})} \right)_{\rm hadron~1} -
\left( {M_n(F^{\rm t-4})\over M_n(F^{\rm t-2})} \right)_{\rm hadron~2}
\end{equation}
which is, however, still sensitive to the genuine higher-twist corrections,
which depend on the exact specific internal hadron wave function.  We hope that
this argument elucidates the fundamental limitations of the renormalon
approach. Nevertheless, to achieve a better understanding of the renormalon
contributions it is important to study as many different cases as possible. On
the basis of these results we hope to develop a physical interpretation which
explains why this phenomenological approach is successful in some cases,
but fails in others \cite{akhou96}.

So far the renormalon model was used to calculate the power corrections for the
non-singlet part of the structure functions $F_L$ \cite{SMMS96} and $g_1$
\cite{MMMSS96}. The physically equivalent approach based on dispersion
relations of \cite{DMW95,BraunBeneke} was used to calculate the higher twist
contribution to the non-singlet part of the structure functions $F_2$ and $F_3$ \cite{DW96}. 

In the case of the structure function $F_2(x)$ an experimental analysis of
higher twist corrections exists for proton and deuteron targets \cite{VM92}.
Assuming $Q^2$ to be low enough, i.e., below the charm threshold, one obtains
\begin{eqnarray}
C_d & = &\frac{1}{F_2^{\rm d,t-2}} ( \frac{2}{9} P_{\rm S} \otimes F_2^{\rm
0,t-2} + \frac{1}{18} P_{\rm NS} \otimes F_2^{\rm 8,t-2} + P_{\rm G} \otimes G)
\nonumber \\
C_{(p,n)}& = & \frac{1}{F_2^{\rm (p,n),t-2}} [ \frac{2}{9} P_{\rm S} \otimes
F_2^{\rm 0,t-2} + \frac{1}{6} P_{\rm NS} \otimes (\pm F_2^{\rm 3,t-2} + 
\frac{1}{3} F_2^{\rm 8,t-2}) + P_{\rm G} \otimes G ] \, ,
\label{renf2}
\end{eqnarray}
where $F_2^{\rm i,t-2}$, i = $0,3,8$ denote corresponding SU(3) combinations of
parton densities, and $G$ is the gluon twist-2 structure function of the
nucleon. The corresponding expression for the $F_3(x)$ structure function reads
\begin{equation}
C_3 = \frac{1}{F_3^{\rm t-2}} (P_{\rm 3} \otimes F_3^{\rm t-2}) \, .
\label{renf3}
\end{equation}
As already mentioned, so far only the coefficients $P_{\rm NS}$ and $P_3$ are
known.  For a first try, we set in (\ref{renf2}) the coefficient $P_{\rm G}$ to
zero, and approximate $P_{\rm S} \sim P_{\rm NS}$, which results in
\begin{eqnarray}
C_d & = & \frac{1}{F_2^{\rm d,t-2}} (P_{\rm NS} \otimes F_2^{\rm d,t-2})
\nonumber \\
C_{(p,n)} & = & \frac{1}{F_2^{\rm (p,n),t-2}} (P_{\rm NS} \otimes 
F_2^{\rm (p,n),t-2}) \, .
\label{renf2a}
\end{eqnarray}

Equations (\ref{renf2}), (\ref{renf3}), and (\ref{renf2a}) define the
renormalon model of higher twist corrections. In the comparison of the model
predictions with the data we have used the NMC analysis of $F_2(x,Q^2)$
\cite{NMC92} for $C_{2,{\rm proton}}$ and $C_{2,{\rm deuteron}}$ as well as the
more precise one for $C_{2, proton} - C_{2,neutron}$.  For $F_3$ we take the
new fit of \cite{Si96}.  We have also confronted our results with the output of
the computer code kindly provided to us by B. Webber \cite{DW96}, but with the
LO GRV \cite{GRV94} parametrization consistently used to represent twist-2
structure functions.

\begin{figure}
\psfig{figure=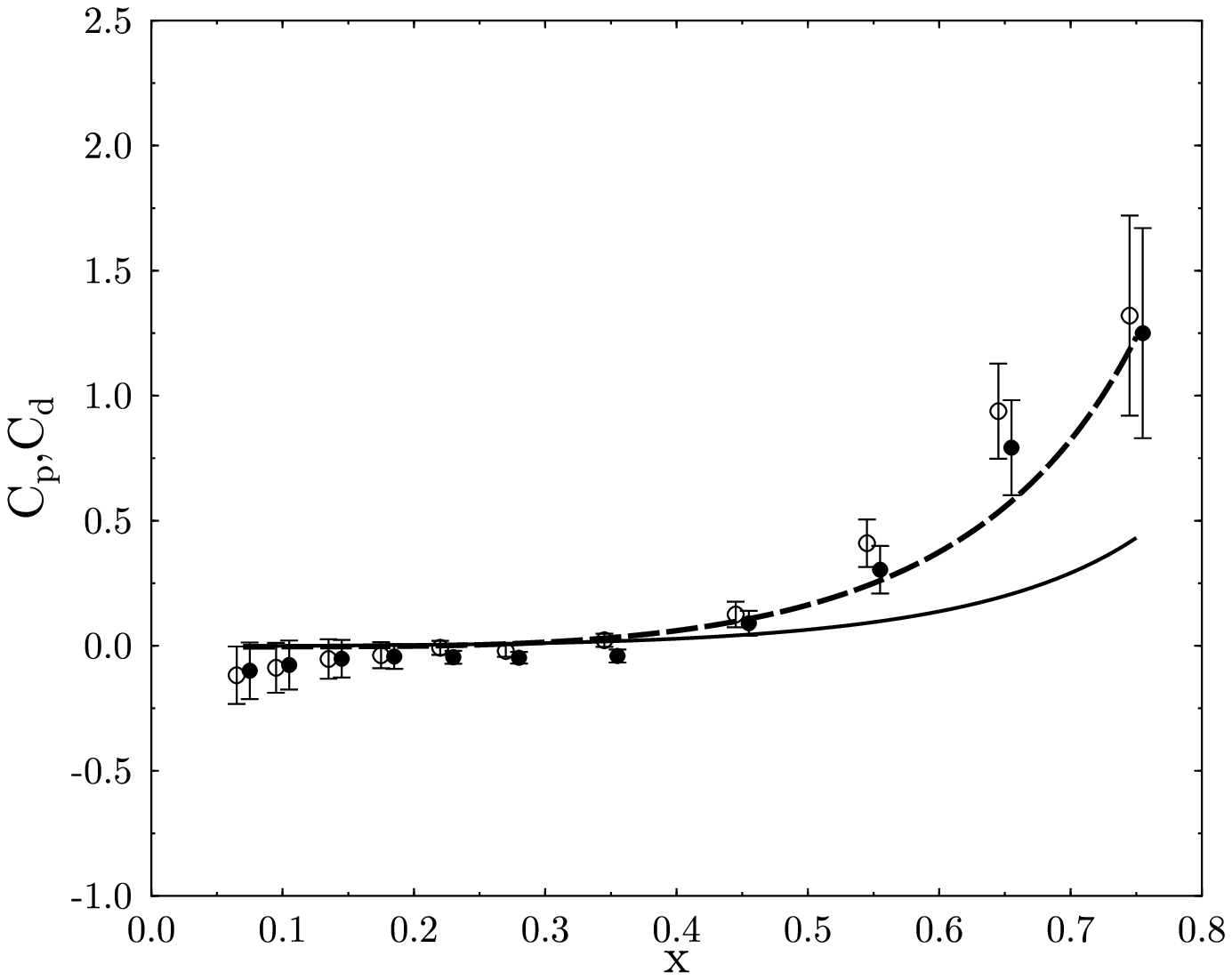,height=4in}
\caption{Renormalon contribution to the coefficient $C_p$ and $C_d$  according
  to Eq. (\ref{renf2a}).  The solid line shows the renormalon model predictions
  for $C_p$ and $C_d$, which are indistinguishable, obtained with the LO GRV
  parametrization \cite{GRV94}. The dashed line shows the fit, as in
  Ref.~\cite{DW96}, which corresponds to the same $x$ dependence, but the value
  of the scale parameter $\mu^2$ about 3 times larger than the one which
  follows from the renormalon model.  The filled and empty circles display the
  data for $C_p$ and $C_d$ according to Ref.~\cite{VM92}, respectively. }
\label{fig1}
\end{figure}

Fig. \ref{fig1} shows our results for $C_p(x)$ and $C_d(x)$ of $F_2$, with
twist-2 structure functions parametrized according to the leading order
GRV-parametrization \cite{GRV94}, as well as the experimental results.
Obviously the renormalon model underestimates the experimental data. On the
other hand the proton and deuteron data differ only slightly, indicating the
dominance of the flavor singlet contribution.

The $x$-dependence of the higher twist correction can be, however, reproduced
very well if one adopts the value of $\mu^2$ a factor of 3 larger 
than what follows
from the renormalon model, $\mu^2 \approx 3 \mu^2_R$, as it was done in 
Ref.~\cite{DW96}. Strictly speaking, the authors of \cite{DW96} used twist-2 MRS(A)
\cite{MRSA} parton distributions normalized at the scale $Q^2 = 10$ GeV$^2$ and
hence they obtained $\mu^2 \approx 2.4 \mu^2_R$. Note that as this scale
dependence is governed by anomalous dimensions of twist-2 rather than twist-4
operators, it differs from what one would expect in QCD.

\begin{figure}
\psfig{figure=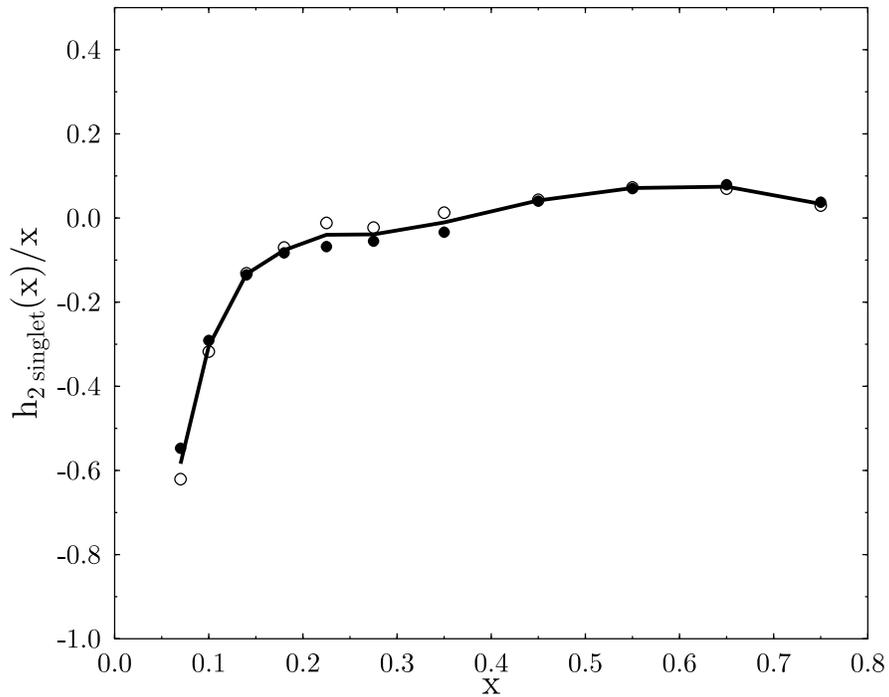,height=4in}
\caption{Determination of the unknown flavor-singlet component of the
  renormalon model from $C_p$ (filled circles) and $C_d$ (empty circles), see
  Eqs.  (\ref{singlet1}) and (\ref{singlet2}) respectively. The LO GRV
  parametrizations \cite{GRV94} normalized at $Q^2 = 2$ GeV$^2$ have been used
  to generate the twist-2 parton distributions. 
  The solid line is the arithmetic
  average.}
\label{fig2}
\end{figure}

\begin{figure}
\psfig{figure=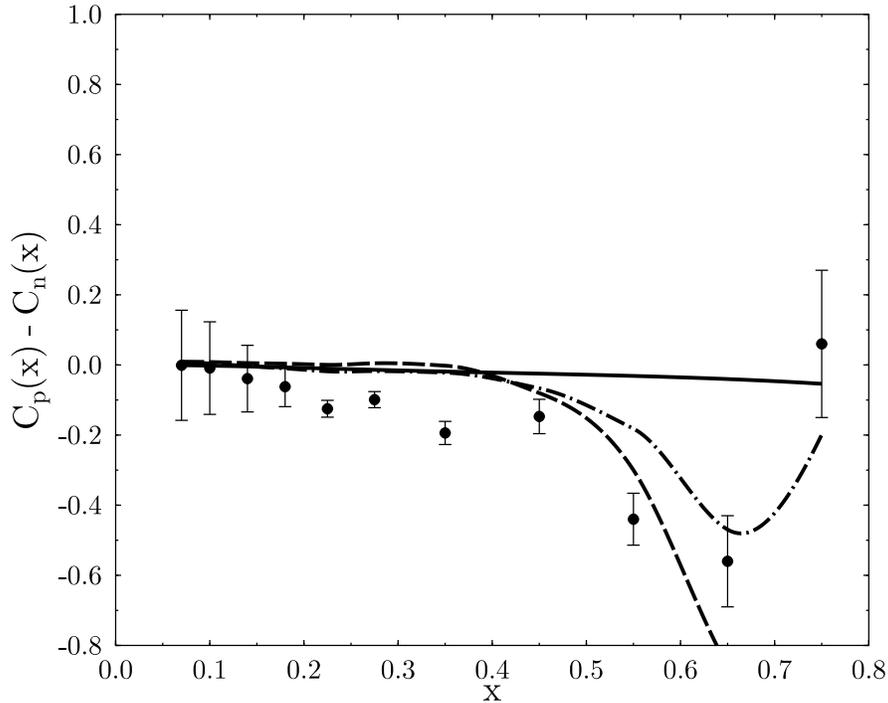,height=4in}
\caption{The renormalon model prediction for $C_p-C_n$ using the LO GRV
  parametrization \cite{GRV94}, solid line, where $C_p$ and $C_d$ have been
  calculated according to Eq. (\ref{renf2a}). Note that due to its definition,
  see Eq. (\ref{30}), $C_p-C_n$, is not a pure non-singlet quantity. The dashed
  line shows the prediction for $C_p-C_n$ after the unknown flavor-singlet
  contribution has been adjusted to reproduce the data for $C_p$ and $C_d$, see
  Eqs.  (\ref{singlet1}) and (\ref{singlet2}).  The dashed-dotted line shows
  the result of the same procedure, but with the scale $\mu^2$ adjusted, 
  like in
  \cite{DW96}, to the description of the coefficients $C_p$ and $C_d$, see
  Figure (\ref{fig1}). The data points are from Ref.~\cite{NMC92}.}
\label{fig3}
\end{figure}

It is interesting to speculate how taking into account the yet unknown
coefficients $P_{\rm S}$ and $P_{\rm G}$ may influence the renormalon model
predictions. First, note that the characteristic rise of the renormalon model
prediction in Figure \ref{fig1}, which clearly follows the trend seen in the
data, results from the term in $P_{\rm NS}$ proportional to $N$ in the Mellin
space \cite{MMMSS96,DW96}. If one assumes that $P_{\rm S}$ and $P_{\rm G}$
follow the same pattern, i.e. they do not contain terms proportional e.g. to
$N^2$, then the gluonic contribution should be always negligible at large $x$.
Indeed, the twist-2 gluon distribution should fall at least one power of $1-x$
faster than the quark one \cite{Bro95}, and therefore it cannot be responsible
for the discrepancy observed in Figure \ref{fig1}. One could then try to
constrain the singlet coefficient $P_{\rm S}$ taking into account the
experimental fact that $C_d(x) \sim C_p(x)$, and known relations between
various parton distributions when $x \to 1$ \cite{Tho95}, but it leads to the
relation $P_{\rm S} = P_{\rm NS}$, which has been discussed already above.
Finally, we note that the deuteron and proton contributions to
Eq. (\ref{renf2}) contain one unknown combination $\frac{2}{9}P_{\rm S} \otimes
F_2^{\rm 0,t-2} + P_{\rm G} \otimes G$ which therefore can be determined
independently from $C_p$ and $C_d$ data:
\begin{equation}
\frac{2}{9} P_{\rm S} \otimes F_2^{\rm 0,t-2} + P_{\rm G} \otimes G  =  
C_d \, F_2^{\rm d,t-2} - \frac{1}{18} P_{\rm NS} \otimes F_2^{\rm 8,t-2} \, ,
\label{singlet1}
\end{equation}
and
\begin{equation}
\frac{2}{9} P_{\rm S} \otimes F_2^{\rm 0,t-2} + P_{\rm G} \otimes G  =  
C_p \, F_2^{\rm p,t-2} - \frac{1}{6} P_{\rm NS} \otimes (F_2^{\rm 3,t-2} + 
\frac{1}{3} F_2^{\rm 8,t-2}) \, .
\label{singlet2}
\end{equation}
We have explicitly checked that both equations (\ref{singlet1}) and
(\ref{singlet2}) are indeed consistent with each other, see Figure \ref{fig2},
and so one can take e.g. the arithmetic average between both estimates as the
approximation for the unknown flavor-singlet contribution. Note that the curve
is negative for $x \le 0.35$ in accordance with the fact that the renormalon
model prediction for $C_p$ and $C_d$ is higher than the data in this region,
see Figure \ref{fig1}. It turns out, however, that when one wants to use this
approximation for comparison with the difference $C_{2,proton}-C_{2,neutron}$,
which has been extracted in \cite{NMC92} with much smaller experimental errors,
one encounters a problem.  The results are shown in Figure \ref{fig3} together
with the now fixed renormalon model prediction (including the fitted quark
singlet and gluonic parts).  The agreement with the data becomes better, but
the model still underestimates the higher twist corrections for $x$ between 0.2
and 0.4.

\begin{figure}
\psfig{figure=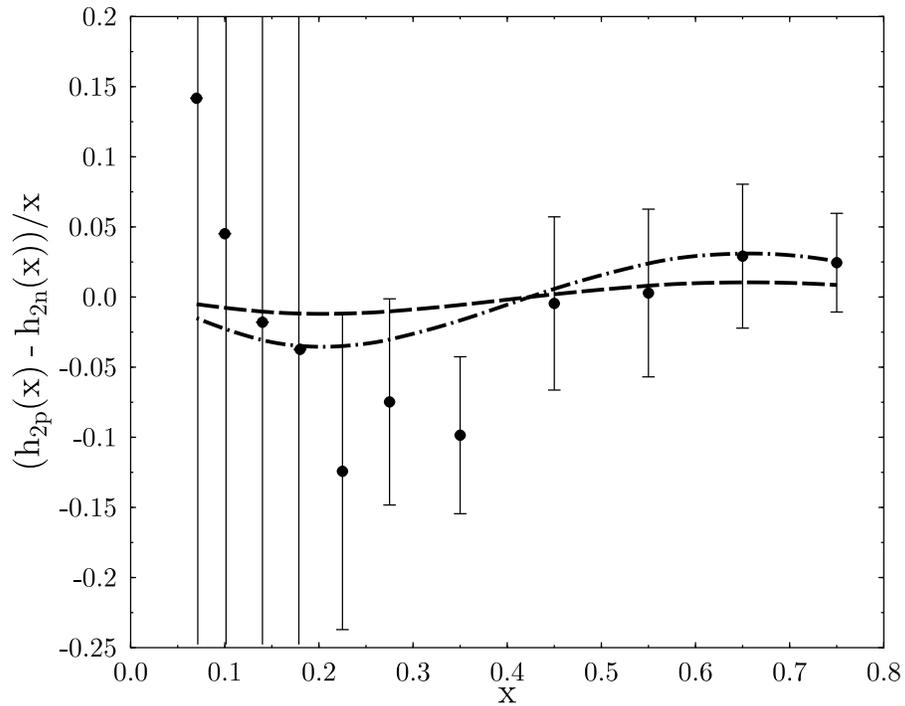,height=4in}
\caption{ The renormalon model prediction for $(h_{2p}(x) - h_{2n}(x))/x$
   using the LO GRV
  parametrization \cite{GRV94}, dashed line.  The dot-dashed line shows the
  prediction with the scale $\mu^2$ adjusted to the description of the
  coefficients $C_p$ and $C_d$, as in \cite{DW96}. The data points are from
  Ref.~\cite{VM92}. Error bars for data points lying below $x = 0.2$ have
  been cut to fit into the figure.}
\label{fig4}
\end{figure}

In addition we have compared the renormalon model prediction for the difference
of coefficients $(h_{2p}(x) - h_{2n}(x))/x$ as given in Ref.~\cite{VM92}, 
see Figure
\ref{fig4}. As far as the $x$-dependence is concerned, the general trend seems
to be reproduced by the model, but the scale has again to be taken a
factor of 3 or larger than the $\mu^2_R$. We note that as 
$(h_{2p}(x) - h_{2n}(x))/x$ is
free from flavor-singlet contributions, it is an ideal observable for models of
higher-twist corrections, and that one should try to get better data (smaller
error bars) for this quantity.

\begin{figure}
\psfig{figure=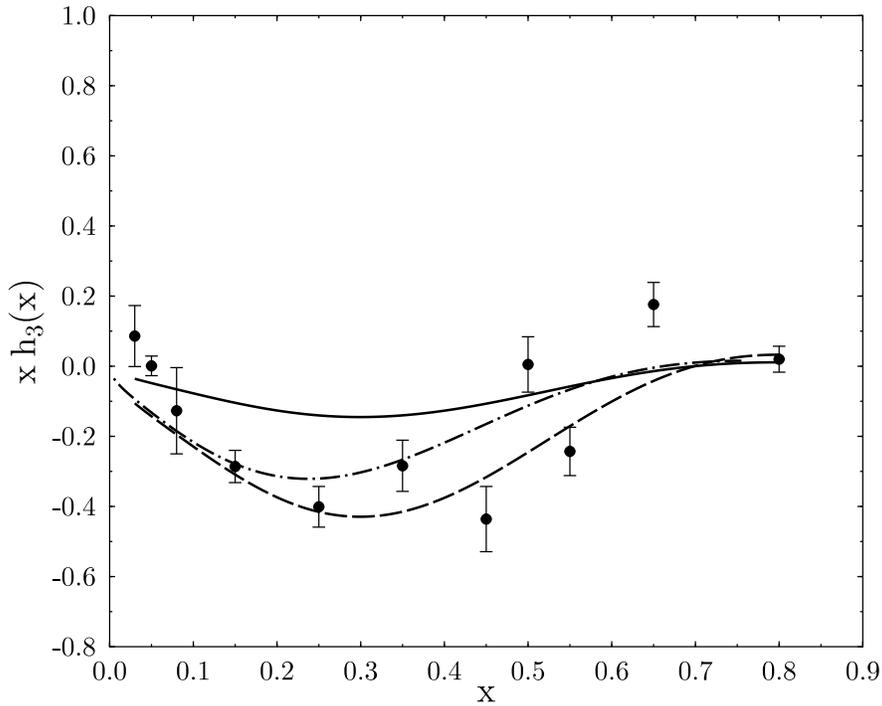,height=4in}
\caption{Renormalon prediction for $x h_3(x)$ using the LO GRV \cite{GRV94} 
  parametrization (solid line). The data points taken from Ref.~\cite{Si96}
  correspond to the LO analysis. The dashed line shows the prediction with the
  scale $\mu^2$ adjusted to the description of the coefficients $C_p$ and
  $C_d$, as in \cite{DW96}. The dot-dashed line shows the original prediction
  of \cite{DW96}, obtained using the MRSA parametrization \cite{MRSA}
  normalized at $Q^2 = 10$ GeV$^2$.}
\label{fig5}
\end{figure}

\begin{figure}
\psfig{figure=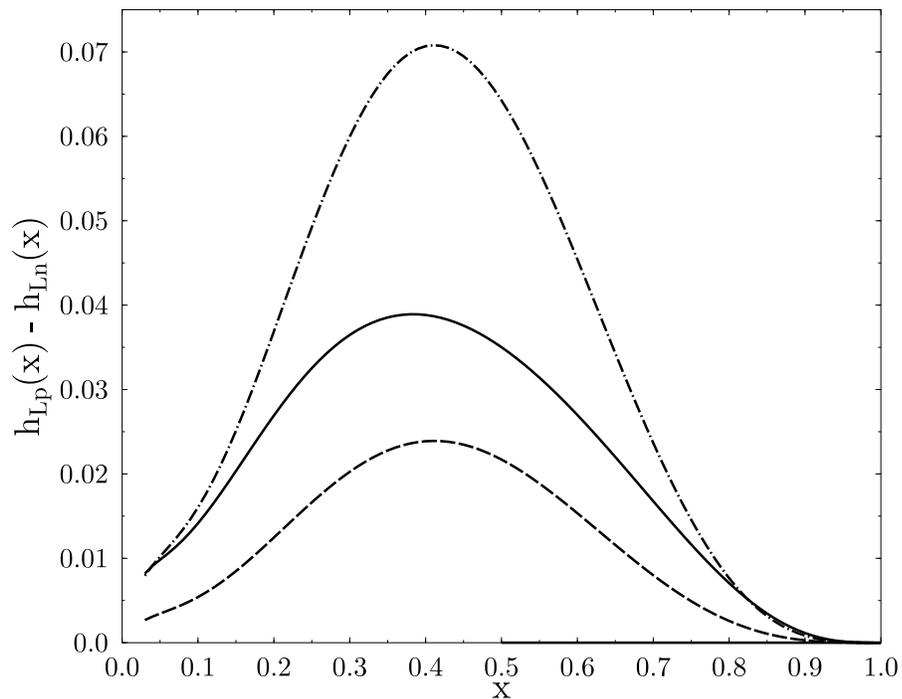,height=4in}
\caption{Coefficient $h_{Lp}-h_{Ln}$, obtained from the phenomenological
  parametrizations to $R(x,Q^2)$ \cite{WRBDR90} and $F_2$ \cite{NMCt4}
  (solid line). 
  The renormalon model
  prediction for $h_{Lp}-h_{Ln}$ using the LO GRV parametrizations is 
  depicted by the
  dashed line. The dot-dashed line shows the prediction with the scale $\mu^2$
  adjusted to the description of the coefficients $C_p$ and $C_d$, as in
  \cite{DW96}.}
\label{fig6}
\end{figure}

The case of the purely non-singlet structure function $F_3(x)$ is illustrated on
Figure \ref{fig5}. Again, although the renormalon model prediction seems to
have the correct shape, it falls short by a factor of 2 or 3 with respect to
the magnitude of the higher-twist effects, as required by the data according to
the LO analysis of \cite{Si96}. Note that while the analysis of the magnitude
of the coefficient $x h_3(x)$ in \cite{Si96} shows that it slightly decreases
from LO to NNLO fits, its characteristic shape, indicating an increase from a
negative to a positive value somewhere at $0.2 \le x \le 0.65$, remains stable.
Finally, an analysis along the lines of Ref.~\cite{SMMS96} shows that the
renormalon model underestimates the magnitude of higher-twists corrections to
$F_L$ by a factor of 2, see figure \ref{fig6}.  Adjusting the scale $\mu^2$ to
the higher twist corrections to $F_2(x,Q^2)$, as in \cite{DW96}, results this
time in a prediction which is larger than what follows from the present data.

Summarizing, we conclude that the model in its present form can certainly be
considered as a useful phenomenological guide for estimates of higher-twist
contributions to DIS, but one has still to be careful in claiming a
phenomenological success of this approach. Our analysis shows that although the
renormalon model does in fact a good job as far as the $x$-dependence of the
higher-twist correction is concerned, the mass scale it predicts is by a factor
2 - 3 too low, depending on the observable. In other words, the NNA estimates
of the leading powered-suppressed corrections to DIS turn out to be a factor
2 or 3 smaller then what seems to be required by the data. Other estimates, for
example the results presented in \cite{MMMSS96}, should be seen from the same
perspective.  Another possibility would be to fit the mass scale independently
to each observable, a point of view which has been recently advocated in
\cite{BBM97}.  One should also keep in mind that the analysis of $F_2(x)$ data
still reveals some discrepancies and it is difficult to say now whether taking
into account the singlet part, possibly with its specific mass scale, will
improve the situation. We hope that these observations will help to understand
successes and limitations of the renormalon phenomenology. 
\\ \\ 
{\bf Acknowledgment:} The authors are grateful to A. Br\"ull and B. Webber for
many useful discussions. This work has been supported by BMBF, DFG (G.~Hess
Programm), DESY, and German-Polish exchange program X081.91.  A.S.~thanks the
MPI f\"ur Kernphysik in Heidelberg for support. L.~M.~was supported in part by
the KBN grant 2~P03B~065~10.
%


\newpage

\end{document}